\documentclass[]{aa}
\usepackage{txfonts}
\usepackage[dvips]{graphicx}
\usepackage{subfigure}

\usepackage{epsfig}

\def\bfv2    {\bf}   

\def\lax    {\ifmmode{_<\atop^{\sim}}\else{${_<\atop^{\sim}}$}\fi}
\def\gax    {\ifmmode{_>\atop^{\sim}}\else{${_>\atop^{\sim}}$}\fi}
\def\kms    {\ifmmode{{\rm ~km~s}^{-1}}\else{~km~s$^{-1}$}\fi}

\def\approx   {$\sim$}
\def\arcm   {$^{\prime}$}
\def\arcmper  {\ifmmode \rlap.{' }\else $\rlap{.}' $\fi}

\def\arcs   {$^{\prime\prime}$}
\def\arcsper  {\ifmmode \rlap.{'' }\else $\rlap{.}'' $\fi}
\def\arcsgper  {\ifmmode \rlap.^{s }\else $\rlap{.}^s $\fi}

\def\deg      {\ifmmode^\circ\else$^\circ$\fi}     

\def\hper     {\ifmmode \rlap.^{h}\else $\rlap{.}^h$\fi}

\def\m1       {$^{-1}$}

\def\mper     {\ifmmode \buildrel m\over . \else $\buildrel m\over .$\fi}

\def\solmass  {M$_\odot$}
\def\sper     {\ifmmode \rlap.^{s}\else $\rlap{.}^s$\fi}

\def\to       {\rightdysonarrow}

\def\>           {$>$}
\def\<           {$<$}

\def\simlt       {\lower.5ex\hbox{$\; \buildrel < \over \sim \;$}}
\def\simgt       {\lower.5ex\hbox{$\; \buildrel > \over \sim \;$}}

\def\apjs   {{\it ApJS{\rm,}\ }}

\def\aap    {{\it A{\rm\&}A{\rm,}\ }}

\begin{document}
\title{The large asymmetric HI envelope of the isolated 
galaxy NGC~864 (CIG 96)}

\author{Espada, D.\inst{1}
\and
Bosma, A.\inst{2}
\and
Verdes-Montenegro, L.\inst{1}
\and 
Athanassoula, E.\inst{2}
\and 
Leon, S.\inst{1}
\and
Sulentic, J.\inst{3}
\and
Yun, M. S.\inst{4}}

\offprints{daniel@iaa.es}
\institute{Instituto de Astrof\'{\i}sica de Andaluc\'{\i}a, CSIC,
Apdo. 3004, 18080
Granada, Spain
\and
Observatoire de Marseille, 2 Place le Verrier, 13248 Marseille Cedex 4, 
France. 
\and
Department of Astronomy, University of Alabama, AL 35487, USA
\and
Department of Astronomy, University of Massachussetts, Amherst, MA 01003, USA
}

\date{Received ; accepted }

\authorrunning{Espada et al.}

\abstract{We present a
HI synthesis imaging study of NGC~864 (CIG~96),
a spiral galaxy well isolated from similarly
sized companions, yet presenting an intriguing asymmetry in its integral
HI spectrum. 
The asymmetry in the 
HI profile is associated with a strong kinematical perturbation in
the gaseous envelope of the galaxy, where at one side the decay of the 
rotation curve is faster than
keplerian.
We detect a small 
(M(HI) = 5 $\times$ $10^{6}$ \solmass )
galaxy with a faint optical counterpart  at $\sim$ 80 kpc 
projected distance from NGC~864.
{ This galaxy is probably not massive enough to have caused the
perturbations in NGC~864. We discuss alternatives,
such as the accretion of a gaseous companion at a radial velocity
lower than the maximum.}
}

\titlerunning{The large asymmetric HI envelope of NGC~864}

\maketitle

\section{Introduction}
The origin of asymmetries  
{ in 
isolated} galaxies is not well understood.
Are they always induced
by the presence of small companions? What is the influence of recent
captures, 
and how long does it take for a parent galaxy to become again
axisymmetric afterwards? Could some originate from internal, as yet
not well studied, long-lived dynamical instabilities  
(e.g. Baldwin et al. 1980)?
Integral HI spectra provide a powerful statistical
approach to this issue since they contain information about both the
HI density distribution and the velocity field.
We have observed, 
or compiled from the literature, single-dish 21-cm line spectra for
almost 800 galaxies (Espada et
al. in prep.) as part of the  AMIGA project (Analysis
of the interstellar Medium of Isolated GAlaxies; 
Verdes-Montenegro et al. 2005), which aims to 
provide a baseline against which environmental effects 
can be evaluated. We have constructed a complete sample of the most
isolated galaxies, starting from the Catalogue of Isolated Galaxies
(CIG, Karachentseva 1973; 1050 galaxies), selected
on the basis of the distance to the nearest similarly sized
galaxies.
About 50\%
of these galaxies show asymmetric HI profiles, consistent
with the results from
previous work on smaller 
samples of isolated galaxies (104 galaxies in Haynes et al. 1998; 
30
galaxies in Matthews et al. 1998).  Surprisingly, other
samples of galaxies
in denser environments show only slightly larger values of 
this rate (50 - 80\%, e.g.  Swaters et al. 2002, Richter \& Sancisi
1994, Sulentic \& Arp 1983). 

HI synthesis imaging of very isolated galaxies
with a highly lopsided 
{ HI profile} 
may clarify the
origin of these asymmetries. Not only HI-rich companions can be
identified, but the presence of tidal features can be revealed,
which would trace a past interaction.
We have selected from our database a well defined
sample of the most isolated galaxies showing significant asymmetries
in their HI profiles (A$ _{n}$ \>
1.1, see Haynes et al. 1998 for the definition), 
and mapped them with the 
VLA\footnote{The National Radio Astronomy Observatory is a facility of
the National Science Fundation Operated under cooperative agreement by
Associated Universities, Inc.} in the 21-cm HI line
in order to 
look for signs 
{ of external}
interaction by analyzing in detail the HI distribution and its
kinematics. 


One of the most interesting galaxies in our 
VLA sample is NGC~864,
{ number 96 in the CIG catalog,
which, as we shall see,}
shows a peculiar HI envelope 
in our 21-cm maps.
It is
classified as SAB(rs)c by de Vaucouleurs et al. (1991). It has
an optical size of    4\arcm.7 x   3\arcmper 5 and an apparent 
magnitude m$_B$ = 11.6. 
{ It has a very asymmetric HI profile 
(cf. Fig. \ref{fig:comparison}).}
 The systemic heliocentric
velocity that we have measured for this galaxy (see Sec.~\ref{sec:model}) is 
1561.6 \kms, 
corresponding to a distance of 17.3 Mpc 
(H$_0$ = 75 km s$^{-1}$ Mpc$^{-1}$) 
{ after applying the correction w.r.t. the CMB reference frame.}
This gives an optical luminosity of 1.1 $\times$ 10$^{10}$ L$_{\odot}$. 
The closest galaxy to NGC~864 at the same redshift is  
 a small companion that we detected in our 
HI observations (see Sec.~\ref{sec:obs})
 with v$_{hel}$ = 1605 \kms, located  at a distance
 d $\sim$ 80 kpc (15\arcmper 3) 
of NGC~864, visible in the POSS2 as a faint object (identified in 
NED as 2MASX J02162657+0556038 with a reported magnitude of  m$_{B}$ =
16.38)
with an approximate size of 0\arcmper8 $\times$ 0\arcmper6 (physical diameter
 $\sim$ 4 kpc). The other galaxies within a projected radius of d = 0.5 Mpc
from NGC~864  are fainter by  3 to 7 magnitudes.


\section{HI synthesis observations and results}
\label{sec:obs}

Observations of the 21 cm line for NGC~864 were made with the VLA in
its D configuration in July 2004 with 26 antennas.  A
bandwidth per IF of 3.125 MHz (from 1249.5 to 1895.2 km s$^{-1}$) was
used, in 2IF correlator mode, giving a velocity resolution of 48.8
kHz (10.4 km s$^{-1}$) for the 64 individual channels after
Hanning smoothing. The data were calibrated using the standard VLA
 procedure in AIPS and were imaged using IMAGR. The
average of the line-free channels has been subtracted from all the
individual channels.  The galaxy was detected between 1436.7 km s$^{-1}$
and 1686.6 km s$^{-1}$. 
The synthesized beam was 49\arcsper 8 $\times$ 46\arcsper 2
($\alpha \times \delta$). 
{ The rms noise level achieved after 4 hours is 0.66 mJy/beam. 
 Primary beam correction has been applied to our maps. In Fig.~\ref{fig:ch-smo-center-a}~.a) and Fig.~\ref{fig:ch-smo-center-b}~.b) we display the channel maps smoothed to a beam size of 70\arcsper 4
$\times$ 65\arcsper 3, containing the HI emission at the observed radial velocities, from 1426.3 to 1707.4\kms.}  
We  detect the small companion in the HI map from { 1572.0 to 1655.3\kms} 
 at { $\alpha$(2000.0)} = 02$^{h}$16$^{m}$26\arcsper9 and { $\delta$(2000.0)} = 05$^{\circ}$56\arcm24\arcsper0 .

The total spectrum has been obtained by integrating the emission in the 
individual
channel maps (solid line in Fig.~\ref{fig:comparison}, top).
We have measured a HI content of M$_{HI}$ = 7.53
$\times$ 10$^9$ M$_{\odot}$,
in very good agreement with   the 
single dish profile obtained by Haynes et
al. (1998, dotted line in Fig.~\ref{fig:comparison}, top),
showing that there is no loss of flux in the synthesis imaging.
 The profile
of the small companion is shown at the bottom of
the same figure.  Subtraction of the flux of
this dwarf galaxy from
 the total spectrum will not change its (asymmetric) shape.


In Fig.~\ref{fig:mom0-all}  
we show the integrated HI column density
distribution of NGC~864 (left) and 
the HI velocity field (right), 
both overlaid on a POSS2 red band image, 
 the deepest one we found. 
These maps 
have been calculated as follows. The channel maps were smoothed
with a gaussian tapering function, 
and then
non-signal pixels of the maps were blanked out. The   
channel maps where emission was detected
were added up to produce the integrated emission map.
The intensity weighted mean radial velocity field has been
obtained in a similar way.
The small companion mentioned 
above 
is clearly seen, together with its faint 
optical counterpart.

The integrated emission map shows  an unresolved 
HI depression in the center
of the galaxy, while
the highest column densities are 
located in a pseudoring structure with a size of 
$\sim$ 90\arcs\ $\times$ 55\arcs , better seen in
the grey scale map shown in 
 Fig.~\ref{fig:mom0-all-grey}.
This seems to trace the outer spiral structure of NGC~864, although the
spatial resolution of the HI data is too small for a detailed comparison.
An arm-like feature seems to join this structure at a second 
enhancement in the column density distribution. This
enhancement has a 
{ ring-like shape}
with an approximate size of 8\arcmper2 $\times$ 4\arcmper 6 
(Fig.~\ref{fig:mom0-all} { and~\ref{fig:mom0-all-grey}}),
nearly doubling the size of  the optical disk.
It is more prominent NW and SE of the optical disk and slightly closed to the 
NE, while
undetected to the SW.
The receding side of NGC~864 is narrower
than the approaching one.
 The velocity field shows a symmetric pattern of
differential rotation in the inner
parts (r $\le$  300\arcs ), while the kinematics of the 
outer parts is strikingly different.
The outer isovelocity contours in the northern part show some
twisting in the E direction, the last contour being nearly closed
as a sign of a flattening rotation curve, and the southern
isovelocity contours bend to the south with decreasing values 
characteristic of a declining rotation curve.
The optical disk is reasonably symmetric in both spiral structure and 
angular extent,
{ except for a slight enhancement of the southern arm, and
a minor extension of the optical disk to the SE.}


\section{Modeling of the galaxy}
\label{sec:model}

We have modeled the velocity-field of NGC~864 with a least-square 
algorithm based on a tilted ring model 
(Begeman 1987; the ROTCUR task in { GIPSY}).
 By this procedure 
we divided the galaxy into concentric rings, each of them
 with a width of 20\arcs\ along the
major axis.
 The asymmetry of the HI distribution and velocity field
 in the outer parts 
 lead us to model separately  the approaching and the receding part of 
the galaxy. Points within a
 sector of $\pm$ 30$^{\circ}$ from the minor axis 
 were excluded from the fits.
{ The center position was fixed to the position of the optical center
$\alpha$(2000.0) = 02$^{h}$15$^{m}$27\arcsper6 { and } 
$\delta$({ 2000.0}) =  06$^{\circ}$00\arcm 09\arcsper1 (Leon \& Verdes-Montenegro 2003).} 
The systemic velocity of 1561.6 km s$^{-1}$ has been obtained
as the central velocity
of the HI spectrum at the 20\% level. 
Expansion velocities were set to zero.
In a first iteration 
the position angle{,  the inclination} and
rotation velocity have been left free,
giving already a very satisfactory 
modeling of the data cube. 
The derived rotation curve,  however, suffers from beam smoothing: 
since the beam is
 elongated along the { major} axis the integration effect biases the
 radial velocities towards values lower than the true
 values corresponding to the rotation curve. 
A second iteration has been performed, this time fixing the 
radial velocities at slightly higher values in order to correct
for beam smearing, and leaving again the position angle  { and the inclination free}.
The  modeled cubes  for the redshifted and
blueshifted parts of NGC~864 were combined in a single cube from which 
a velocity field has been obtained ({ Fig.~\ref{fig:model}}).
It shows a remarkable agreement with the observed one 
(Fig.~\ref{fig:mom0-all} { right}), although not 
reproducing the asymmetries along the minor axis direction.
Deriving the position angle of NGC~864 from our modeling
for r $\le$ 300\arcs\ we obtain a mean value of 
23$^{\circ}$ $\pm$ 3, in very good agreement with the optical disk
orientation. { The inclination obtained as the mean between the receding and approaching sides is 43$^{\circ}$ $\pm$ 2, which is consistent to the value given by Tully (1988) of 45$^{\circ}$. The modelled inclinations for both sides are almost constant and their values are always close to the mean}. The major axis twists both in the northern and southern part
of NGC~864, although is more pronounced to the north. 
There it starts at r $\sim$
240\arcs\ at nearly the edge of the optical disk, reaching a position
angle of 39$^{\circ}$. In the southern part it starts at  r $\sim$
360\arcs\ going up to 31$^{\circ}$.
In Fig.~\ref{fig:posvel} we compare the position-velocity cut 
at 23$^{\circ}$
for the modeled (top) and observed (bottom) cube.
Again the observations are very well reproduced. The plot also
shows the { projected rotation} curve used. 
This figure contains the essential peculiarities of NGC~864:
asymmetric HI distribution and kinematics, with a noticeable
drop in rotation velocities. 
We note that the blueshifted part of the 
observed position-velocity diagram shows a
faster than keplerian drop in the velocity of NGC~864, 
with a trend to reach r$^{-0.5}$ values at larger
radii, { that should of course be explained by non-circular motions and/or projection effects}. Such a phenomenon is extremely rare in HI envelopes.

\section{Discussion and conclusions}

Our results for NGC~864
have wider implications :\\

a) The integral HI profile is symmetric in velocity, but asymmetric
in intensity. Yet the 2D-kinematics of NGC~864 show large scale 
asymmetries:  there is a kinematic warp, and
in the southern part there is an abrupt decline of the rotation
velocity, as is evident in Fig.~\ref{fig:posvel}.
The atomic gas here looks like a kinematically detached clump {, }
evident as a secondary peak in the position velocity cut at
radii $\sim$ 400\arcs\ to the { SW}. { This region has a physical extent of around 3\arcm$\times$4\arcm ~ in the major axis and minor axis directions, respectively.}
Clearly a simple integrated profile analysis will conclude
that the central parts of NGC~864 are roughly symmetric, with a
value of $\Delta$V$_{20}$ correlated with the luminosity
of the galaxy, since it falls on the Tully-Fisher
relation. It is the outer parts which are surprisingly asymmetric,
but since the radial velocities in the approaching side
 are closer to the systemic
velocity than those in the central parts, 
{ the asymmetry} only manifests itself
in a higher amplitude of the blueshifted horn of
the HI profile. \\

 b) The outer HI envelope is large, massive, very
asymmetric and { presents much} structure. 
Its HI mass is about 4.5
$\times$ 10$^9$ M$_{\odot}$, and the HI mass
associated with the steep drop of the rotation curve is about 
2.3 $\times$ 10$^8$ M$_{\odot}$. 
No optical counterpart is found associated to the perturbed
atomic gas. Warped HI envelopes are common around spirals, but 
few have so much structure and asymmetries as the one
reported here.\\

 c) The galaxy is isolated with respect to similarly sized galaxies
 by a rather strict criterion. Yet there are 5 
small companions within a projected distance of 500 kpc. 
The closest one,  detected in our maps, has
a { dimensionless gravitational interaction strength} (Dahari 1984) of 
4 $\times$ 10$^{-4}$, while for the others this parameter
ranges from 1 $\times$ 10$^{-5}$ to 8 $\times$ 10$^{-7}$.
They are hence too small in mass
to have caused the strong perturbations in the outer envelope of NGC~864.
A similar conclusion follows from considerations about spiral
forcing by companions (cf. Athanassoula 1984).\\

{ We have considered several possibilities to explain the origin 
of these HI asymmetries.
{ A self-induced perturbation seems highly unlikely: while the pseudoring 
is a very standard feature, the outer HI ring-like structure is too large 
to be a broken resonance ring, since the outer Lindblad resonance is usually 
located a bit over 2 times the end of the bar (Athanassoula et al 1982). An 
external perturbation by a companion, which could have interacted or even 
been accreted, seems likely to have caused the {SW} clump and the 
outer ring-like structure. 
However, we can exclude an encounter
with a large pericenter distance since this would necessitate a
massive companion, which is simply not there. A small companion would
have to come very near NGC 864 or, even better, go through it, but a
central, or near-central, passage will (cf.
Athanassoula et al. 1997, Berentzen et al. 2003) yield
results that do not match at all the
morphology of NGC 864. The alternative left is that the companion
crossed the equatorial plane of the target at an intermediate
distance, e.g. just outside the optical disc and still within
the extended HI disc. Such a passage could have induced the warp,
and if the intruder was a loosely bound gas rich dwarf, its gas
could have contributed to the SW clump, which has a similar gas mass, 
while its stars and dark matter could have dispersed.
To study this scenario, which is of course only a cartoon -- and to 
explain the form of
the ring-like density enhancement, the mass of the excess gas and the
velocity perturbations -- full blown
selfconsistent simulations, including both gas and stars, are necessary,
but beyond the scope of the present paper. }

\begin{acknowledgements}
DE, SL and LVM are partially supported by DGI Grant
AYA 2002-03338
and Junta de Andaluc\'{\i}a TIC-114 (Spain).
LV-M acknowledges the hospitality
of the Observatoire de Marseille. 
{ We used} 
the NASA/IPAC Extragalactic Database (NED), 
operated by the Jet Propulsion Laboratory, California Institute 
of Technology, 
under contract with the National Aeronautics and Space Administration.
\end{acknowledgements}

\begin{figure*}
{\baselineskip=1.0mm \hfill}
\begin{center}  
\caption{Comparison of the integrated VLA profile (solid line)
and the HI profile obtained by Haynes et al. (1998) at 
Green Bank 43m (dots). At the bottom we show the HI profile
of the dwarf companion found in the HI map.}


\label{fig:comparison}
\end{center}
\end{figure*}

\begin{figure*}
\begin{center}  
\caption{HI column density distribution (left) and velocity field (right) of NGC~864  and its companion, superimposed on the optical POSS2 red band image. The contour levels are: 5, 12, 24, 37, 49, 61, 73, 86, 98, 110 and 122 $\times$ $10^{20}$ cm$^{-2}$. The velocity contours go from 1465 to { 1660 km s$^{-1}$} in intervals of 15 km s$^{-1}$, and are labeled each 30 km s$^{-1}$. The beam size of 49\arcsper 8 $\times$ 46\arcsper 2 is shown in the upper left  of all panels.}
\label{fig:mom0-all}
\end{center}
\end{figure*}


\begin{figure*}
\caption{Grey scale map of the HI column density distribution of NGC~864.
 { Darker regions} correspond to higher column densities. The beam size of 49\arcsper 8 $\times$ 46\arcsper 2 is shown in the upper 
left part. The  dashed line 
defines the approximate location of the pseudoring described in 
$\S$~\ref{sec:obs}. The insert shows the inner structure of NGC~864.}
\label{fig:mom0-all-grey}
\end{figure*}

\begin{figure*}
\begin{center}  
\caption{ .a) HI channel plots superimposed on the optical POSS2 red band image. The beam size of
 70\arcsper 4 $\times$ 65\arcsper 3 is shown in the upper 
left part. { The contour levels are:  3, 8, 16, 32, 48, 64, 80 and 96 mJy/beam. }} 
\label{fig:ch-smo-center-a}
\end{center}
\end{figure*}
\setcounter{figure}{3}
\begin{figure*}
\begin{center}  
\caption{ .b) Continued.}
\label{fig:ch-smo-center-b}
\end{center}
\end{figure*}


\begin{figure*}
\hfill
\begin{center}  
\caption{{ HI velocity modeled as explained in $\S$~\ref{sec:model}.
The velocity field contours
are plotted in the map in km $s^{-1}$ from
1465 to { 1660 km s$^{-1}$}, in intervals of 15 km s$^{-1}$. }}
\label{fig:model}
\end{center}
\end{figure*}

\begin{figure*}
\hfill
\begin{center}  
\caption{Position velocity along 
p.a. = 23$^{\circ}$ for the model (top),
and for the observations (bottom). The { projected }rotation curve is also 
plotted (dots).}
\label{fig:posvel}
\end{center}
\end{figure*}

\begin{figure*}
\begin{center}  
\caption{{ Rotation curves of NGC~864 considering: a) both the receding and approaching regions, b) the approaching half (southern region) and c) the receding half only (northern region). The dashed line represents the rotation curve in a) without error bars. } }
\label{fig:compara-rot}
\end{center}
\end{figure*}


\begin{thebibliography}{}
\bibitem{} Athanassoula, E. 1984, Physics Report, 114, 321
\bibitem{} Athanassoula, E., Bosma, A., Cr\'ez\'e, M. \& Schwarz, M.P.
           1982, A\&A 107, 101
\bibitem{} Athanassoula, E., Puerari, I., Bosma, A. 1997, MNRAS 286, 284
\bibitem{} Baldwin, J. E., Lynden-Bell, D. \& Sancisi, R. 1980, MNRAS 193, 313
\bibitem{} Begeman, K. 1987, Ph. D. Thesis, University of Groningen
\bibitem{} Berentzen, I., Athanassoula, E., Heller, C.H. \& Fricke, K.J.
           2003, MNRAS, 341, 343
\bibitem{} Dahari, O. 1984, AJ 89, 966
\bibitem{} De Vaucouleurs, G.,  De Vaucouleurs, A., Corwin, H. G., et al. 
           1991, Third
            Reference Catalog of Bright Galaxies (Springer, Berlin) 
           AJ 95, 697. 
\bibitem{} Haynes, M. P., van Zee, L., Hogg, D. E., Roberts, M. S., Maddalena, R. J. 1998 AJ 115, 62	
\bibitem{}  Karachentseva, V. E. 1973, Comm. Spec. Ap. Obs., USSR 8, 1
\bibitem{} Leon, S. \& Verdes-Montenegro, L. 2003, A\&A 411, 391
 \bibitem{} Matthews, L. D., van Driel, W., Gallagher, J. S., III 1998 AJ 116 1169	
\bibitem{} Richter, O. G. \& Sancisi, R., 1994, A\&A 290, 9
\bibitem{} Sulentic, J. W. \& Arp, H. 1983, AJ 88, 489
\bibitem{} Swaters, R. A., van Albada, T. S., van der Hulst, J. M. \& Sancisi,
           R. 2002, A\&A 390, 829
\bibitem{} Tully, R. B. 1988, Nearby Galaxies Catalog, Cambridge Univ. Press, 
           Cambridge
\bibitem{} Verdes-Montenegro, L., Sulentic, J., Lisenfeld, U.,
           Leon, S., Espada, D., Garcia, E., Sabater, J. \& Verley, S. 2005, 
           A\&A (in press)
\end{thebibliography}
\end{document}